%% file: mainLong.tex
\documentclass[journal,twocolumn]{IEEEtran}
\input{headers.tex}
\begin{document}
\pdfminorversion=4
\title{Control of Asynchronous Imitation Dynamics on Networks}
\author{James Riehl$^\dagger$, Pouria Ramazi$^\dagger$ and Ming Cao$^\dagger$}
\maketitle
\thispagestyle{empty}
\pagestyle{empty}

\renewcommand{\thefootnote}{*} \footnotetext[1]{This work was supported in part by the European Research Council (ERCStG-307207).}

\renewcommand{\thefootnote}{$\dagger$}
\footnotetext[2]{ENTEG, Faculty of Mathematics and Natural Sciences, University of Groningen, The Netherlands, \{j.r.riehl,p.ramazi,m.cao\}@rug.nl}

\maketitle
\thispagestyle{empty}
\pagestyle{empty}

\begin{abstract}
Imitation is widely observed in populations of decision-making agents. Using our recent convergence results for asynchronous imitation dynamics on networks, we consider how such networks can be efficiently driven to a desired equilibrium state by offering payoff incentives for using a certain strategy, either uniformly or targeted to individuals. In particular, if for each available strategy, agents playing that strategy receive maximum payoff when their neighbors play that same strategy, we show that providing incentives to agents in a network that is at equilibrium will result in convergence to a \textit{unique} new equilibrium. For the case when a uniform incentive can be offered to all agents, this result allows the computation of the optimal incentive using a binary search algorithm. When incentives can be targeted to individual agents, we propose an algorithm to select which agents should be chosen based on iteratively maximizing a ratio of the number of agents who adopt the desired strategy to the payoff incentive required to get those agents to do so. Simulations demonstrate that the proposed algorithm computes near-optimal targeted payoff incentives for a range of networks and payoff distributions in coordination games. 
\end{abstract}

\section{Introduction}
Networks in which agents make decisions by imitating their most successful neighbors appear frequently in sociology, biology, economics, and engineering \cite{traulsen2010human,van2015focus}.
Such networks of \textit{success-based} learners often exhibit complex non-convergent behaviors even when the agents are homogeneous.
In particular, these networks are less likely to converge than networks of agents who use best responses; in other words, focusing on the success of others hinders the agents from reaching satisfactory decisions \cite{ramazi2017imitation}.
This non-convergence relates to volatility and instability, which can have consequences ranging from costly inefficiencies to catastrophic failures.
Imitation is also known to lead to selfish behaviors in various social contexts \cite{van2015focus}, which can manifest as social dilemmas such as tragedy of the commons, in which the pursuit of selfish goals leads to globally suboptimal outcomes.
However, in many of these cases it may be possible to circumvent the undesired global outcomes by administering some small control input to the agents, locally.
Given that this could require a large amount of total control effort, it is critical to develop methods for achieving these goals as efficiently as possible. 
Game theory is widely used to model distributed optimization and learning in large populations of autonomous agents \cite{cortes2015self,marden2014generalized,li2014decoupling,wikecek2011stochastic,shamma2005dynamic,guo2013algebraic,gharesifard2012evolution,mylvaganam2015constructive}, but more specifically, \emph{evolutionary game theory} allows for strategies to propagate through populations by means other than rational choice, and therefore provides an ideal framework to model networks of imitative agents \cite{cheng2015modeling,ramazi2016networks,nowak2006evolutionary,zhu2016evolutionary,ramazi2016evolutionary,ramazi2015feeling}.
The use of payoff incentives to drive populations towards a desired strategy is gaining in popularity \cite{montanari2010spread,zhu2016evolutionary,JamesMing2,ramazi2015analysis}. 
Several approaches have been used to control such networks, including offering incentives uniformly to the agents in a network \cite{Liang2015,riehlControl}, targeting individual agents with incentives \cite{riehlControl}, and directly controlling the strategies of some of the agents \cite{james,JamesMing2}. 
There are two key properties that facilitate the design of control algorithms for each of these cases; if the network is at some equilibrium, then providing incentives to the agents should (i) cause no agent to switch away from a desired strategy, and (ii) result in convergence of the network to a unique equilibrium state.
In \cite{riehlControl}, we have established these properties for certain types of agents who update asynchronously with best responses, but the conditions under which networks of imitative agents can be driven to a desired equilibrium remain to be discovered.

In this paper, we design efficient incentive-based control algorithms for three different control problems on finite networks of heterogeneous decision-making individuals who asynchronously imitate their highest earning neighbors.
We start by building a general framework for asynchronous network games with two available strategies, $A$ and $B$.
Our main theoretical contribution is to show that in any such network game, regardless of the update rule, if all agents are \textit{$A$-coordinating}, i.e., agents who update to strategy $A$ would also do so if they had more neighbors playing $A$, then providing incentives to the agents when the network is at equilibrium (i) causes no agent to switch from $A$ to $B$, and (ii) leads the network to a unique equilibrium regardless of the agents' activation sequence.  
Next we establish that networks governed by imitation dynamics satisfy these conditions provided that all agents are \textit{opponent coordinating}, i.e., agents' payoffs are maximized when their neighbors play the same strategy that they do.
These results make possible the design of efficient control algorithms, using payoff incentives, to guarantee the convergence of networks of imitative agents to a desired strategy.
First, for the case when an incentive is offered uniformly to all agents, we provide a simple binary search algorithm to compute the optimal incentive. 
Next, when incentives can vary from agent to agent, inspired by our approach for controlling best-response networks \cite{riehlControl}, we propose the \emph{Iterative Potential-to-Reward Optimization (IPRO)} algorithm, which selects agents to be targeted based on iteratively maximizing a weighted ratio of the number of agents who adopt the desired strategy to the payoff incentive required to get those agents to do so. 
Simulations show that the IPRO algorithm achieves near optimal performance in a variety of cases and outperforms several other incentive targeting algorithms based on degree, earnings, and other criteria. 

\section{Asynchronous network games}

Although this paper focuses primarily on imitative agents, since some of the results apply to a broader class of dynamics, we present here a generalized framework for two-strategy asynchronous games on networks.

Consider an undirected network $\G = (\V,\E)$, in which the nodes $\V=\{1,\ldots,n\}$ represent agents and the edges $\E \subseteq \V\times\V$ define two-player games between pairs of adjacent neighbors.
Each agent starts by playing one of the strategies $A$ or $B$ against all neighbors, then accumulates the resulting payoffs, and the process repeats at each time $k\in\{0,1,\ldots\}$.
The possible payoffs to an agent $i\in\V$ are defined by the \emph{payoff matrix} $\pi^i\in\mathbb{R}^{2\times 2}$, whose entry $\pi_{x_i,x_j}$ represents agent $i$'s payoff when playing strategy $x_i$ against a neighbor $j$ playing strategy $x_j$, where $x_i,x_j\in\{A,B\}$, and we define $A=1$, $B=2$ for the purposes of matrix indexing.
For compactness of notation, we stack all payoff matrices in a 3-dimensional matrix $\pi\in(\mathbb{R}^{2\times 2})^n$.
The payoff of agent $i$ against all neighbors at time $k$ is given by
\begin{equation*}
	u_i(k) = \sum_{j\in\N_i}   \pi^i_{x_i(k),x_j(k)},
\end{equation*}
where $\N_i$ is the set of agent $i$'s neighbors. 
After collecting all payoffs, one random agent $i$ activates at time $k$ and updates to a new strategy at time $k+1$ according to an update rule, which we denote by $\Rr$: 
\begin{equation}	\label{updateRule0}
	x_i(k+1) = 
    \begin{cases}
		A & \text{ if } f_i(x(k)) = \{A\}	\\
        B & \text{ if } f_i(x(k)) = \{B\}	\\ 
        z_i & \text{ if } f_i(x(k)) = \{A,B\}
	\end{cases}
\end{equation}
where the function $f_i: \{A,B\}^n\to \{A,B, \{A,B\}\}$. Each $z_i$ is fixed and equals either $A,B$ or $x_i(k)$. 
By a \emph{network game} $\Gamma:(\G,\pi,\Rr)$ we mean a network $\G$ of agents with payoff matrices $\pi$, who update based on the rule $\Rr$. 
We do not prescribe any particular process for driving the activation sequence, but we do make the following assumption, which simply ensures that no agent will stop updating after some amount of time.
\begin{assumption} \label{ass-persistent}
Every agent activates infinitely many times as time goes to infinity. 
\end{assumption}

Agents' strategies evolve under the update rule according to the sequence in which agents activate, and may converge to an equilibrium state or continue to fluctuate.   
An equilibrium of the network game is a state $x^*$ at which no agents will switch strategies under their respective update rules, implying that if $x(k) = x^*$ for some $k\geq 0$, then $x(k+1) = x^*$, regardless of which agent is active at time $k$.
Our eventual goal is to use payoff incentives to control the dynamics of network games such that the network reaches or gets as close as possible to a desired equilibrium state in which every agent plays strategy $A$. 
By offering payoff incentives to a network, we mean offering non-negative rewards for playing $A$ to one or more agents in the network, which equates to adding non-negative constants to the first row of the corresponding payoff matrices.
To set the foundation for the main control results, we first investigate when a network game governed by an arbitrary update rule will reach a unique equilibrium after offering payoff incentives. 

\section{Unique equilibrium convergence of $A$-coordinating network games} \label{sec-uniqEq}

Equilibrium convergence is a key property of network games, and can be guaranteed for certain classes of update rules and agent payoff matrices \cite{ramazi2017imitation,ramazi2016networks}.
However, it is not generally the case that such networks will converge to a \textit{unique} equilibrium, a property which is highly desirable for the design of efficient and predictable control algorithms.
Here we establish conditions on the agents and update rule under which unique equilibrium convergence can be guaranteed.

We say a network game is \emph{$A$-coordinating} if any agent who updates to strategy $A$ would also do so if some agents currently playing $B$ were instead playing $A$. 
Formally, we have the following definition.

\begin{samepage}
\begin{definition}
	A network game $(\G,\pi,\Rr)$ is $A$-coordinating if for any two strategy vectors $y,z\in\{A,B\}^{n}$ satisfying
    \begin{equation}		\label{def1-1}
		y_i = A \Rightarrow z_i = A \qquad \forall i\in\V,
	\end{equation}
 the following holds 
    \begin{align}	
    	f_i(y)=\{A\} \Rightarrow  f_i(z) = \{A\} \qquad \forall i\in\V
		\label{def1-2}
	\end{align}  
and
    \begin{align}	
     	f_i(y) = \{A,B\} \Rightarrow A\in f_i(z) \qquad \forall i\in\V.
		\label{def1-3}
	\end{align}  
\end{definition}
\end{samepage}

The $A$-coordinating property implies that an increase in the number of agents playing $A$ may lead agents to switch from $B$ to $A$ but will never cause agents to switch from $A$ to $B$, yielding a monotone behavior in agents' strategy updates.
We say that a network game is \emph{$A$-monotone} if, after offering payoff incentives to one or more agents when the network is at any equilibrium, no agent will ever switch from $A$ to $B$.
The following proposition holds.
\begin{proposition}	\label{prop-A-monotone}
	Every $A$-coordinating network game is $A$-monotone.
\end{proposition}

We need the following lemma for the proof.
\begin{lemma}		\label{lem-00}
	Consider an $A$-coordinating network game $(\G,\pi,\Rr)$.
    If for some agent $i\in\V$, one of the following holds at some time $k\geq 0$:
    \begin{enumerate}
    	\item $A\in f_i(x(k))$ and  $A\not\in f_i(x(k+1))$, 
        \item $B\not\in f_i(x(k))$ and  $B\in f_i(x(k+1))$,
    \end{enumerate}
    then an agent has switched from $A$ to $B$ at time $k+1$. 
\end{lemma}
\begin{IEEEproof}
	We prove by contradiction. 
    Assume the negation of Lemma \ref{lem-00} holds for a network at some time $k$ and let $x(k)$ denote the state of the network at that time.
    Since no agent has switched from $A$ to $B$ at time $k+1$, the vectors $y = x(k)$ and $z = x(k+1)$ satisfy Condition \eqref{def1-1}.
    Now if Case 1 takes place, then either $f_i(y) = \{A\}$, violating \eqref{def1-2} or $f_i(y) = \{A,B\}$, violating \eqref{def1-3}, a contradiction, yielding the result. 
    If on the other hand, Case 2 takes place, then $f_i(y) = \{A\}$, violating \eqref{def1-2} since $B\in f_i(z)$, a contradiction, leading to the proof. 
\end{IEEEproof}

\begin{IEEEproof}[Proof of Proposition \ref{prop-A-monotone}]
	We again prove by contradiction.
    Assume the contrary and let $k_1\geq1$ be the first time that some agent $i$ switches from $A$ to $B$.
    Then one of the following cases holds: 
    
   	\emph{Case 1:} $A\not\in f_i(k_1-1)$.
    On the other hand, either the strategy of agent $i$ is $A$ at $0$, yielding $A\in f_i(0)$ since the network is at equilibrium at $0$, or there is some time $\hat k\in[0,k_1-2]$ such that agent $i$ switches to $A$ at $\hat k+1$, yielding $A\in f_i(\hat k )$.  
    So in any case, there exists some $k_0\in[0,k_1-1]$ such that $A\in f_i(k_0)$. 
    Therefore, since $A\not\in f_i(k_1-1)$ there exists some time $k_2\in[k_0,k_1-2]$ such that $A\in f_i(k_2)$ and $A\not\in f_i(k_2+1)$.  
    In view of Lemma \ref{lem-00}, this implies that an agent has switched from $A$ to $B$ at $k_2+1$, a contradiction since $k_1>k_2+1$ is the first time that such a switch takes place, yielding the result.
    
    \emph{Case 2:} $ f_i(k_1-1) = \{A,B\}$ and $z_i = B$.
    On the other hand, either the strategy of agent $i$ is $A$ at $0$, yielding $f_i(0) = \{A\}$ since $z_i = B$ and that the network is at equilibrium at $0$, or there is some time $\hat k\in[0,k_1-2]$ such that agent $i$ switches to $A$ at $\hat k+1$, yielding $ f_i(\hat k ) = \{A\}$.  
    So in any case, there exists some $k_0\in[0,k_1-1]$ such that $f_i(k_0)=\{A\}$. 
    Therefore, since $f_i(k_1-1)=\{A,B\}$ there exists some time $k_2\in[k_0,k_1-2]$ such that $B\not\in f_i(k_2)$ and $B\in f_i(k_2+1)$.
    In view of Lemma \ref{lem-00}, this implies that an agent has switched from $A$ to $B$ at $k_2+1$, a contradiction, leading to the proof.
\end{IEEEproof}

Moreover, we say that a network \emph{switches sequence-independently} if, after offering incentives to one or more agents when the network is at equilibrium, any agent who switches from $B$ to $A$ under one activation sequence will do so under any activation sequence (possibly at a different time).
\begin{proposition}	\label{prop-switchSequenceIndependently}
	Every $A$-coordinating network switches sequence-independently.
\end{proposition}

This proposition can be explained intuitively as follows. 
Consider two activation sequences $\str S^1:=\{i^{0}, i^{1}, \ldots \}$ and $\str S^2:=\{j^{0}, j^{1}, \ldots \}$.
Let $i$ be the first agent who switches from $B$ to $A$ under the sequence $S^1$. Since switches from $A$ to $B$ are impossible due to Proposition \ref{prop-A-monotone}, this must be the first switch of any kind. 
The first time agent $i$ is active under the sequence $S^2$, she will also switch from $B$ to $A$ since up to that time, agents may have switched only from $B$ to $A$ under $S^2$, again due to the monotonicity established in Proposition \ref{prop-A-monotone}.
Then by induction, the same can be shown for the second and later agents who switch their strategies from $B$ to $A$ under $S^1$.
We formalize and prove this statement in the following Lemma, borrowing
some ideas from our previous result in \cite{riehlControl}.
Let $t_0$ be the first time when agent $j^{0}$ is active in $\str S^1$.
Then for $s=1,2,\ldots$, define $k_s$ as the first time after $k_{s-1}$ that agent $j^{s}$ is active in $\str S^1$.	
The time $k_s$ exists because of the assumption that each agent activates infinitely many times.
    Denote by $x^1_i$ and $x^2_i$, the strategies of agent $i$ under the activation sequences $\str S^1$ and $x^2_i(k)$, respectively.
\begin{lemma}	\label{lem3}
	Consider an $A$-coordinating network game $(\G,\pi,\Rr)$ which is at equilibrium at time $0$.
    Suppose that some payoff incentives are offered at time $0$.
    Then given any two activation sequences $\str S^1=\{i^{0}, i^{1}, \ldots \}$ and $\str S^2=\{ j^{0}, j^{1}, \ldots \}$, the following holds for $s=0,1,2\ldots$
	\begin{equation}	\label{th1-3}
		x^2_{j^{s}}(s+1) = A 	\quad\Rightarrow\quad x^1_{j^{s}}(k_s+1) = A. 
	\end{equation} 
\end{lemma}
\begin{IEEEproof}
	We prove by induction on $s$.
 	The statement is first shown for $s=0$.
	Suppose $x^2_{j^{0}}(1) = A$.
    The initial strategy of agent $j^0$ is the same under both sequences, i.e.,  $x^1_{j^0}(0) = x^2_{j^0}(0)$.
    Therefore, since the network game is $A$-monotone in view of Proposition \ref{prop-A-monotone}, no agent has switched to $B$ before time $k_0$, under $\S^2$. 
    So since the network game is $A$-coordinating, it follows that $x^1_{j^0}(k_0+1) = A$ if $x^1_{j^0}(1) = A$, verifying \eqref{th1-3} for $s=0$. 
    
	Now assume that \eqref{th1-3} holds for $s=0,1,\ldots,r-1$.
	Suppose $x^2_{j^{r}}(r+1) = A$.
	Now since \eqref{th1-3} holds for all $s=0,1,\ldots,r-1$, and because of Proposition \ref{prop-A-monotone}, we obtain that if any agents have switched and hence fixed their strategies from $B$ to $A$ under $\S^2$ up to the time $k = r$, they have also done so under $\S^1$ up to any time $k \geq k_{r-1}+1$.
    Moreover, no agent has switched from $B$ to $A$ under $\S^1$. 
    Thus, the strategy vectors $y = x^2(r)$ and $z = x^1(k_{r})$ satisfy the condition in \eqref{def1-2}. 
    Hence, \eqref{th1-3} is true for $s = r$ since the network game is $A$-coordinating.
\end{IEEEproof} 
	
\begin{IEEEproof}[Proof of Proposition \ref{prop-switchSequenceIndependently}]
	The proof follows directly from Lemma \ref{lem3}.
\end{IEEEproof}

These two properties of $A$-coordinating network games lead to the main result of this section.
We say that a network game is \emph{uniquely convergent}, if after offering some payoff incentives when the network is at equilibrium, the network will again reach an equilibrium state which is unique and does not depend on the sequence in which agents activate.
\begin{theorem}		\label{th1}
	Every $A$-coordinating network game is uniquely convergent.
\end{theorem}
\begin{IEEEproof}
	According to Proposition \ref{prop-A-monotone}, no agent switches from $B$ to $A$. 
    Since by Assumption \ref{ass-persistent}, every agent activates infinitely many times, it follows that the network will reach an equilibrium state at some finite time after a maximum of $n$ total strategy switches. 
	It remains to prove the uniqueness of the equilibrium for all activation sequences, which we do by contradiction.
	Assume that there exist two activation sequences $\str S^1=\{i^{0}, i^{1}, \ldots \}$ and $\str S^2=\{j^{0}, j^{1}, \ldots \}$ that drive the network to two distinct equilibrium states.
    This implies the existence of an agent $q$ whose strategy differs between the two equilibria, say $B$ under the equilibrium of $\str S^1$ and $A$ under the equilibrium of $\str S^2$, without loss of generality. 
    However, in view of Proposition \ref{prop-switchSequenceIndependently}, agent $q$ will switch from $B$ to $A$ at some time under $\S^1$ and will not change afterwards because of Proposition \ref{prop-A-monotone}, implying that the network was not at equilibrium, which is a contradiction and completes the proof. 
\end{IEEEproof}

\section{Imitation update rule}

The {imitation update rule} $\I$ dictates that agent $i$, active at time $k$, updates at time $k+1$ to the strategy of the agent earning the highest payoff at time $k$ in the neighborhood $\N_i\cup\{i\}$. 
If neighbors using both strategies earn the highest payoff, we assume agent $i$ does not switch:
\begin{equation}		\label{updateRule}
	x_i(k+1) =
	\begin{cases}
		A					&  \str S^M_i(k) = \{A\}	\\
		B					&  \str S^M_i(k) = \{B\}	\\        
		x_i(k)				&  \str S^M_i(k) = \{A,B\}
	\end{cases}
\end{equation}
where $\str S^M_i(k)$ is the set of strategies earning the maximum payoff in the neighborhood of agent $i$, that is
\begin{equation*}
	\str S^M_i(k) \dfb \left\{x_j(k)\,\Big|\, u_j(k) = \max_{r\in\N_i\cup\{i\}} u_r(k) \right\}.
\end{equation*}



Asynchronous imitation updates may not in general result in convergence to an equilibrium, but we have established in \cite{ramazi2017imitation} that such networks will converge when all agents are \emph{opponent coordinating}, i.e., earn maximum payoff when their neighbors play the same strategy that they do.
Equivalently, each diagonal entry of the payoff matrix of an opponent-coordinating agent $i$ is greater than the off-diagonal in the same row:   
\begin{equation}	\label{row-coordination}
	\pi^i_{1,1} > \pi^i_{1,2}, \qquad
    \pi^i_{2,2} > \pi^i_{2,1}.
\end{equation}
The following proposition implies that such networks are also $A$-coordinating. 
\begin{proposition}
	Every network of opponent-coordinating agents who update according to the imitation rule $\I$ is $A$-coordinating.
\end{proposition}
\begin{IEEEproof}
	Consider two strategy vectors $y,z\in\{A,B\}^n$ satisfying \eqref{def1-1}, and let the network be at state $y$.
    First we look at the case when $f_i(y) = \{A\}$ for some agent $i\in\V$, implying that the highest-earning agent in the neighborhood $\N_i\cup \{i\}$ of agent $i$ is an $A$-playing agent. 
    Now, if the strategy of some of the $B$-playing agents are changed to $A$ so that the network reaches $z$, then the payoff of no $A$-playing agent decreases and the payoff of no $B$-playing agent increases since all agents are opponent coordinating. 
    Hence, the highest-earning agent in the neighborhood of agent $i$ will still be an $A$-playing agent, yielding $f_i(z) = \{A\}$, resulting in \eqref{def1-2}.
    The case when $f_i(y) = \{A,B \}$ can be proven similarly.
\end{IEEEproof}
The next corollary follows directly from Theorem \ref{th1}.
\begin{corollary}		\label{cor}
	Every network of opponent-coordinating agents is $A$-monotone and uniquely convergent.
\end{corollary}

That is, when a network of opponent-coordinating agents is at equilibrium at time $0$, if non-negative rewards are offered to one or more agents whenever they play $A$ at some time $k\geq0$, then no agent will switch from $A$ to $B$ at any time $k\geq 1$, and the whole network will reach a unique equilibrium at some finite time.

\section{Control through Payoff Incentives}

Having established the unique equilibrium convergence in networks of opponent-coordinating agents, we now investigate the efficient use of payoff incentives to drive such networks of imitative agents from any undesired equilibrium toward a desired equilibrium in which all or at least more agents play strategy $A$.  

\subsection{Uniform Reward Control}

Suppose that some central agency has the ability to offer a reward of $r_0\geq 0$ to all agents whenever they play strategy $A$. The resulting payoff matrix is given by
\begin{align*} 
\hat \pi^i := \bordermatrix{& A & B \cr
                  A & a_i+r_0 & b_i+r_0   \cr
                  B & c_i & d_i  \cr}, \qquad a_i,b_i,c_i,d_i \in\mathbb{R},
\end{align*}
for each agent $i\in\V$.
Let $\mathbf{B}$ denote the $n$-dimensional strategy vector in which each agent plays $B$. 
The control objective in this case is the following.
\begin{problem}[Uniform reward control] \label{probUniform}
	Given a network game  $\Gamma = (\G,\pi,\I)$ and initial strategies $x(0)\neq \mathbf{B}$, find the infimum reward $r_0^*$ such that for every $r_0 > r_0^*$, every agent will eventually play $A$.
\end{problem}

In networks of opponent-coordinating agents, it is relatively straightforward to compute the optimal value of $r_0^*$ once we have established the properties in Section \ref{sec-uniqEq}. 
We take advantage of two key properties established in Corollary \ref{th1}. 
First, the number of agents who converge to $A$ is monotone in the value of $r_0$ due to the $A$-monotone property.
Second, simulations of the network game are fast to compute due to the unique convergence property.
That is, to test the effect of a particular payoff incentive, since all activation sequences will result in the same equilibrium, we can choose a sequence consisting only of agents who will switch from $B$ to $A$, which will have a maximum length of $n$ before reaching equilibrium. 

We begin by generating a set $\R$ containing all possible candidate infimum rewards. 
This set is generated by computing all possible payoff differences between agents playing $B$ and agents playing $A$ when they are neighbors or linked by another initially $B$-playing agent.
Consider a network of opponent-coordinating agents that is at equilibrium at time zero. 
Let $n^A_i$ denote the number of neighbors of agent $i$ who initially play $A$.
Since no agent will switch from $A$ to $B$, the possible payoffs of an agent $i$ when playing $A$ (resp. $B$) at any time step are contained in the sets
\begin{align*}
  \Pi^A_i &\dfb \left\{a_i(n^A_i + \delta_i) + b_i(\deg_i - n^A_i - \delta_i)
    : \delta_i \in \Delta_i \right\} \\
  \Pi^B_i &\dfb \left\{c_i(n^A_i+\delta_i) + d_i(\deg_i - n^A_i - \delta_i)
    : \delta_i \in \Delta_i \right\},
\end{align*}
where $\Delta_i = \{0,1,\dots,\deg_i-n^A_i\}$.

Now consider an agent $s$ who initially plays $B$ and has a neighbor $j$ whose strategy was either initially $A$ or became $A$ at some other time. 
In order for agent $j$ to cause agent $s$ to switch to $A$, the payoff of agent $j$ must be greater than that of each $B$-playing agent (denoted by $i$) in the neighborhood of agent $s$. 
Therefore, the reward given to agent $j$ must be greater than $\frac{y^{B}_i-y^A_j}{\deg_j}$ for some $y_i^B\in\Pi^B_i$ and $y_j^A\in\Pi^A_j$.
This leads to the following set of all candidate infimum rewards, formally derived in the proof of Proposition \ref{rinR}.
\begin{align*}
	&\R \dfb	
    	\Big\{
        	\frac{y_i^B - y_j^A}{\deg_j} \Big| y_i^B\in\Pi_i^B, y_j^A\in \Pi_j^A,\\
            	&\qquad\qquad\ \ j\in \N_s, i\in\N_s\cup\{s\}, x_i(0) = B, \\
               &\qquad\qquad\qquad\qquad\qquad\quad s\in\V, x_s(0) = B
        \Big\}  \cup \{0\}.
\end{align*} 

\begin{proposition} \label{rinR}
	For a network of opponent-coordinating agents with initial strategies $x(0)\neq \mathbf{B}$, $r^*_0 \in \R$.
\end{proposition}
\begin{IEEEproof}
 	Should all agents' strategies be initially $A$, the result is trivial since $r^*_0 =0$.
 	So consider the situation where at least one $B$-playing agent exists. 
    We observe that the network will reach the state of all $A$ after offering the reward $r$ at time $k=0$, if the following condition is satisfied:
    for every agent $s\in\V$ who initially plays $B$, there exists some time $k^s$ such that
    	$x_s(k^s) = B$ and $x_s(k^s+1) = A$.
    Equivalently, for every initially $B$-playing agent $s\in\V$, there must exist some time $k^s$ and $A$-playing neighbor $j\in\N_s$, $x_j(k^s) = A,$ such that for all $B$-playing agents $i\in\N_s\cup\{s\}$, $x_i(k^s) = B$,
    \begin{equation*}
    	r\deg_j + u_j(k^s) > u_i(k^s). 
    \end{equation*}
    Since $u_j(k^s)\in\Pi^A_j$, $u_i(k^s)\in\Pi^B_i$ and $x_i(0) = B$, the condition is satisfied if the following holds: for every initially $B$-playing agent $s\in\V$, there exists some time $k^s$ and agent $j\in\N_s,$ such that for all $i\in\N_s\cup\{s\}$, $x_i(0) = B$,
    \begin{equation*}
    	r\deg_j + y_j^A > y_i^B 
    \end{equation*}
    for some $y_j^A\in \Pi_j^A$ and some $y_i^B\in\Pi_i^B$.
    Now since this is a sufficient condition for $r$ to drive the network to the all-$A$ state, we have that
    \begin{align*}
    	r^*_0 = 
        \inf&\Big\{
        	r\Big| r > \frac{y_i^B - y_j^A}{\deg_j}, y_i^B\in\Pi_i^B, y_j^A\in \Pi_j^A,\\
            	&\qquad\qquad\ \ j\in \N_s, i\in\N_s\cup\{s\}, x_i(0) = B, \\
               &\qquad\qquad\qquad\qquad\qquad\quad s\in\V, x_s(0) = B
        \Big\},
    \end{align*}
    implying that
    \begin{align*}
    	r^*_0 \in 
        &\ \Big\{
        	r\Big| r = \frac{y_i^B - y_j^A}{\deg_j}, y_i^B\in\Pi_i^B, y_j^A\in \Pi_j^A,\\
            	&\qquad\qquad\ \ j\in \N_s, i\in\N_s\cup\{s\}, x_i(0) = B, \\
               &\qquad\qquad\qquad\qquad\qquad\quad s\in\V, x_s(0) = B
        \Big\} \\
        =&\ \R -\{0\}.
    \end{align*}
    By summarizing this case and the case when $r^*_0 = 0$, we complete the proof.
\end{IEEEproof}

Next we sort the elements of $\R$ from low to high and denote this vector by $v^\R$.
Algorithm \ref{algSearchR} performs a binary search over $v^\R$ to find the infimum reward such that all agents in the network will eventually play $A$. 
Denote by $\mathbf{1}$ the $n$-dimensional vector containing all ones. 
In what follows, we also denote by $\bar x$ the unique equilibrium resulting from a particular set of incentives being offered to a network of $A$-coordinating agents starting from $x$.    
\begin{algorithm}[h] \label{algSearchR}
\SetAlgoLined\DontPrintSemicolon
 $i^- := 1$\; 
 $i^+ := |\R|$\;
 \While{$i^+-i^- > 1$}{
   $r_0^* := v^\R_j$, where $j:=\ceil{\frac{i^-+i^+}{2}}$\;
   $\hat \Gamma := (\G,\hat \pi,\I)$\;
   Simulate $\hat\Gamma$\ from $x(0)$ until equilibrium $\bar x$\;
   \uIf{$\bar x_i = A$ for all $i\in\V$}{
     $i^+ := j$\; }
   \Else{$i^- := j$\;}
 }  
\vspace{3pt}
\caption{Binary search over candidate rewards to find the value of $r_0^*$ that solves Problem \ref{probUniform} for networks of opponent-coordinating agents.} 
\end{algorithm}

\begin{samepage}
\begin{proposition}
Algorithm \ref{algSearchR} computes the reward $r_0^*$ that solves Problem \ref{probUniform} for networks of opponent-coordinating agents. 
\end{proposition}
\end{samepage}

\begin{IEEEproof}
Since $r_0^*\in\R$ due to Proposition \ref{rinR}, we can restrict our search to the set $\R$.
Due to Theorem \ref{th1}, we know that if a given incentive $r_0$ results in all agents switching to $A$ for one activation sequence, then it does so for every activation sequence.
Therefore, we can determine the exact equilibrium resulting from offering an incentive $r_0$ by using any activation sequence.
Now consider two incentives $q>p\geq0$.
Corollary \ref{cor} implies that no agent can switch from $B$ to $A$ after offering an incentive to a network at equilibrium.
Hence, if an agent $i$ who initially plays $B$ does not switch to $A$ under incentive $q$, then that agent will also not switch to $A$ under incentive $p$.
Otherwise, offering the additional incentive $q-p$ to a network at the equilibrium  resulting from incentive $p$ would cause this agent to switch from $B$ to $A$, a contradiction.
It follows from this monotonicity property that a binary search on the ordered list $v^\R$ will yield the solution to Problem \ref{probUniform}. 
\end{IEEEproof}


\subsection{Targeted Reward Control}

Suppose that rather than offering a uniform incentive to all agents who play strategy $A$, one has the ability to offer a different reward to each agent.
By targeting the most influential agents in the network, it may be possible to achieve the desired outcome at much lower cost than with uniform rewards, but which agents should be targeted and how much reward should be offered to each of these agents?

Let $r := (r_1,\dots,r_n)^T$ denote the vector of rewards offered to each agent, where $r_i \in\mathbb{R}_{\geq 0}$ is the reward offered to agent i, resulting in the following payoff matrix for each agent $i\in\V$:
\begin{align*} 
\hat \pi ^i := \bordermatrix{& A & B \cr
                  A & a_i+r_i & b_i+r_i   \cr
                  B & c_i & d_i  \cr}, \quad a_i,b_i,c_i,d_i \in\mathbb{R}.
\end{align*}
The targeted control objective is the following. 
\begin{problem}[Targeted reward control] \label{probTargeted}
Given a network game  $\Gamma = (\G,\pi,\I)$ and initial strategies $x(0)\neq \mathbf{B}$, find the targeted reward vector $r^*$ that minimizes $\sum_{i\in\V} r_i^*$ such that if $r_i > r_i^*$ for each $i$, then every agent will eventually play $A$.
\end{problem}

Towards a solution to this problem, we observe that for a network at some equilibrium state $\bar x$, the only way to get imitating agents to switch from $B$ to $A$ through positive rewards is to offer those rewards to agents who start at $A$ or will switch to $A$ at some time and who have at least one neighbor playing $B$. 
For such an agent, the infimum reward such that at least one $B$-playing neighbor will switch to $A$ is  
\begin{equation} \label{minReward}
\check r_i = \max_{j\in \N_i^B} \max_{k\in \N_j^B} \bar u_k - \bar u_i,
\end{equation}
where $\bar u_i$ denotes the payoff of agent $i$ when the state of the network is $\bar x$, and $\N_i^B := \{j\in\N_i \cup \{i\} : \bar x_j = B\}$ denotes the self-inclusive set of neighbors of agent $i$ who are playing $B$.
Due to Corollary \ref{cor}, offering this incentive to agent $i$ will result in a unique equilibrium regardless of the activation sequence.
As a result, we can repeatedly use \eqref{minReward} to construct a targeted reward vector starting from the previous equilibrium.
Indeed, an algorithm which iteratively offers rewards in this manner can be used to compute a reward vector that achieves uniform convergence to $A$.
A generic version of such an algorithm is described below, in which the key step is the choice of the agent $j$ at each iteration, and $\epsilon$ denotes an arbitrarily small positive constant.
\begin{algorithm}[h] \label{algTargeted}
\SetAlgoLined\DontPrintSemicolon
Initialize $\bar x_i = x_i(0)$ and $r_i = 0$ for each $i\in\V$\;
\While{$\exists i\in\V : \bar x_i \neq A$}{
  $\A^B := \{i\in\V:\bar x_i = A \wedge [\exists j\in\N_i :     \bar x_j = B]\}$\; 
  Choose an agent $j\in\A^B$\;
  Let $r_{j} := r_{j} + \check{r}_{j} + \epsilon$\;
  $\hat \Gamma := (\G,\hat \pi,\I)$\;
  Simulate $\hat\Gamma$\ from $\bar x$ until equilibrium $\bar x'$\;
  $\bar x := \bar x'$
}
\vspace{3pt}
\caption{Generic iterative algorithm for computing a reward vector such that all agents in the network will play strategy $A$.} 
\end{algorithm}

It is possible to find the exact solution $r^*$ to Problem \ref{probTargeted} using Algorithm \ref{algTargeted} to perform an exhaustive search on every possible sequence of agents in the initial set $\A^B$.
However, the computational complexity of such an algorithm prohibits its use on large networks.
In Section \ref{sims}, we explore the use of various heuristics for choosing an agent to target at each iteration, including random selection, maximum degree, and maximum payoff earnings.
Next, we propose a slightly more advanced heuristic for incentive targeting inspired by a similar approach to controlling best-response networks \cite{riehlControl}.

Consider a network of opponent-coordinating agents, which is at some equilibrium state $\bar x$. 
In order to identify which agents should be offered incentives, we propose a simple potential function 
\begin{equation}
\Phi(x) = \sum_{i=1}^n n^A_i(x), 
\end{equation}
where $n^A_i(x)$ denotes the number of neighbors of agent $i$ who play strategy $A$ in the state $x$.
This function has a unique maximum, which occurs when all agents play $A$, and increases whenever an agent switches from $B$ to $A$.
Problem \ref{probTargeted} is thus equivalent to the problem of finding the infimum reward vector that maximizes $\Phi(x)$.
Therefore we propose a type of greedy algorithm which iteratively chooses the agent who maximizes the ratio of the change in potential to the reward required to achieve that change.
Let $\bar x'$ denote the equilibrium resulting from offering the reward $\check r_j$ to agent $j$.
We define the iterative potential to reward algorithm (IPRO) as Algorithm \ref{algTargeted} in which the targeted agent is selected as follows.
\begin{equation} \label{minReward}
j^* = \arg\max_{j\in \A^B} \frac{\Delta\Phi(\bar x)^\alpha}{\check r_j^\beta},
\end{equation}
where $\Delta\Phi(\bar x) := \Phi(\bar x') - \Phi(\bar x)$, and $\alpha,\beta \geq 0$ are free design parameters.

\subsection{Budgeted Targeted Reward Control}

In this section, we suppose that there is a limited budget from which to offer rewards and pose the following variation to Problem \ref{probTargeted}.

\begin{problem}[Budgeted targeted reward control] \label{probBudget}
	Given a network game $\Gamma = (\G,\pi,\I)$, initial strategy state $x(0)$, and budget constraint $\sum_{i\in\V} r_i < \rho$, find the reward vector $r^*$ that maximizes the number of agents in the network who will eventually play $A$.
\end{problem}

Algorithm \ref{algBudgetTargeted} slightly modifies Algorithm \ref{algTargeted}
to approximate the solution to Problem \ref{probBudget}. The only difference is that the algorithm will now terminate if no more agents can be offered a reward without violating the budget constraint $\rho$.

\begin{algorithm}[h] \label{algBudgetTargeted}
\SetAlgoLined\DontPrintSemicolon
Initialize $\bar x_i = x_i(0)$ and $r_i = 0$ for each $i\in\V$\;
\While{$\exists i\in\V : \bar x_i \neq A$ and $\displaystyle\sum_{i\in\V} r_i < \rho$}{
  $\A^B := \{i\in\V:\bar x_i = A \wedge [\exists j\in\N_i :     \bar x_j = B]\} \wedge \check{r}_i \leq \rho - \sum_{i\in\V} r_i\}$\;
  Choose an agent $j\in\A^B$\;
  Let $r_j := r_j + \check{r}_j + \epsilon$\;
  $\hat \Gamma := (\G,\hat \pi,\I)$\;
  Simulate $\hat\Gamma$\ from $\bar x$ until equilibrium $\bar x'$\;
  $\bar x := \bar x'$
}
\vspace{3pt}
\caption{Generic iterative algorithm for computing a reward vector to approximate the maximum number of agents who will play strategy $A$.} 
\end{algorithm}




\vspace{-10pt}
\section{Simulations} \label{sims}

Here we compare the performance of the IPRO algorithm to some alternative approaches for controlling networks of agents with imitative dynamics. 
Each of these methods is applied iteratively, targeting agents with payoff rewards until either the control objective is achieved or the budget limit is reached.
Short descriptions of each algorithm under consideration are provided below. 
\begin{itemize}
\item \textbf{Iterative Random (rand):} target random agents in the network
\item \textbf{Iterative Degree-Based (deg):} target agents with maximum degree 
\item \textbf{Iterative Maximum Earning (IME):} target $A$-playing agents earning the highest payoffs while having at least one neighbor playing $B$
\item \textbf{Iterative Potential Optimization (IPO):} target agents resulting in the maximum increase of the potential function ($\alpha=1$, $\beta=0$)
\item \textbf{Iterative Reward Optimization (IRO):} target agents requiring minimum reward ($\alpha=0$, $\beta=1$) 
\item \textbf{Iterative Potential-to-Reward Optimization (IPRO):} target agents maximizing the potential-change-to-reward ratio ($\alpha=1$, $\beta=1$)
\item \textbf{Optimal:} perform exhaustive search to find optimal solution (only practical for small networks)
\end{itemize}
For each set of simulations, we generate geometric random networks by randomly distributing $n$ agents in the unit square and connecting all pairs of agents who lie within a distance $R$ of each other.

Heterogeneous payoffs for the agents are generated as follows: $\pi_i = p_iI + v_iW_i$, where $p_i\geq 1$ denotes the coordination level, $v_i \in [0,1]$ denotes the payoff variance, and $W_i$ is a $2\times 2$ matrix whose elements are drawn independently at random from a uniform distribution on the interval $[0,1]$. Also, the $W_i$ matrices are independent across all agents.
Next, we present four brief simulation studies and provide graphical results, which are also summarized in Table \ref{SummaryTable}.

\subsection{Uniform vs. Targeted Reward Control}

\begin{figure}[h] 
\centering
\includegraphics[width=.8\linewidth]{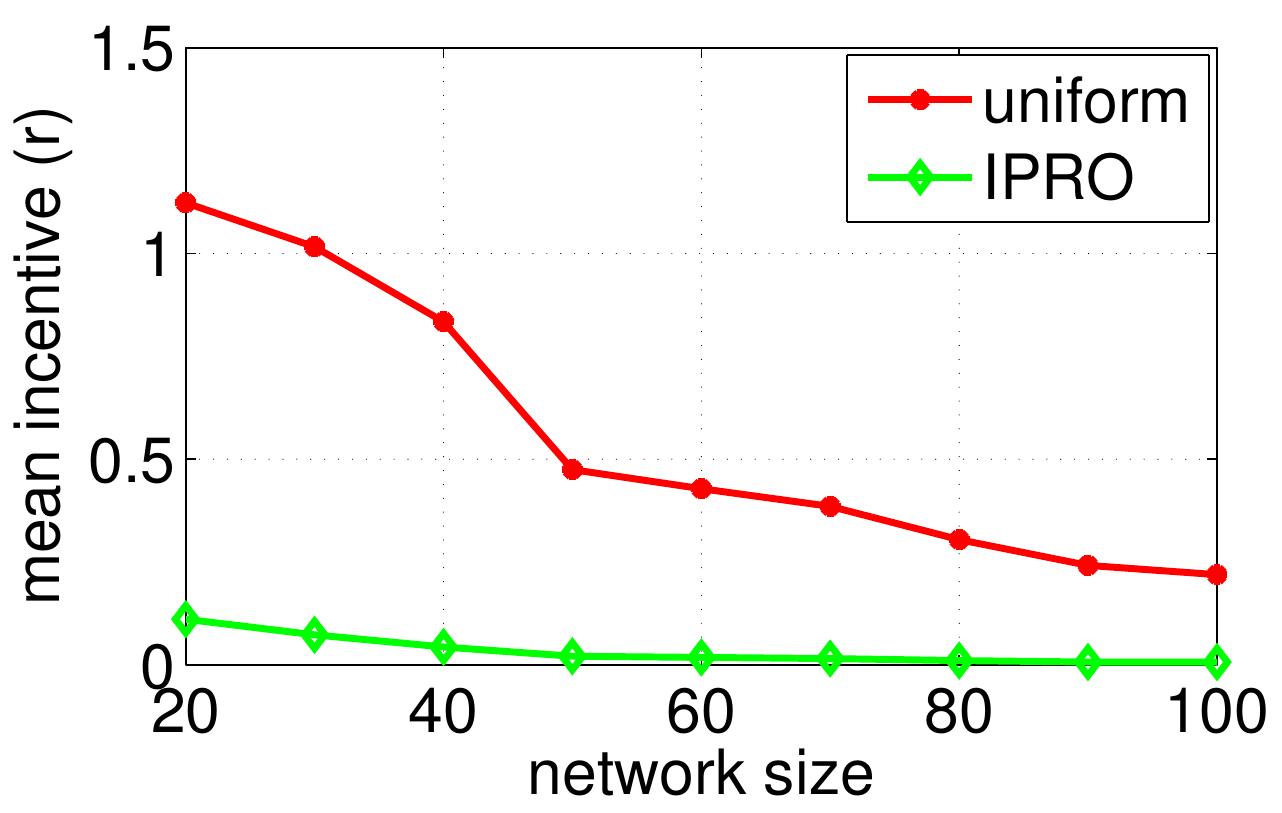}
\caption{Comparison of uniform and targeted reward control on geometric random networks for a range of sizes. For each size tested, 100 random networks were generated using a connection radius $R = \sqrt{(1+\deg_{\text{exp}})/\pi n}$, corresponding to a mean node degree of approximately $\deg_{\text{exp}}=4$. The parameter values used to generate the payoffs were $p_i=1$ and $v_i=\frac{1}{2}$.}
\label{figUniformVsTargeted}
\end{figure}

First, we investigate the difference between uniform and targeted reward control to estimate the expected cost savings when individual agents can be targeted for rewards rather than offering a uniform reward to all agents. 
Fig. \ref{figUniformVsTargeted} shows that targeted reward control offers a large cost savings over uniform rewards, but also that the savings decreases as the networks get larger. Notably, this differs from our findings on best-response networks, in which the opposite effect was observed \cite{riehlControl}. 

\subsection{Targeted-Reward Control: Network Size}

\begin{figure}[h] 
\centering
\includegraphics[width=.8\linewidth]{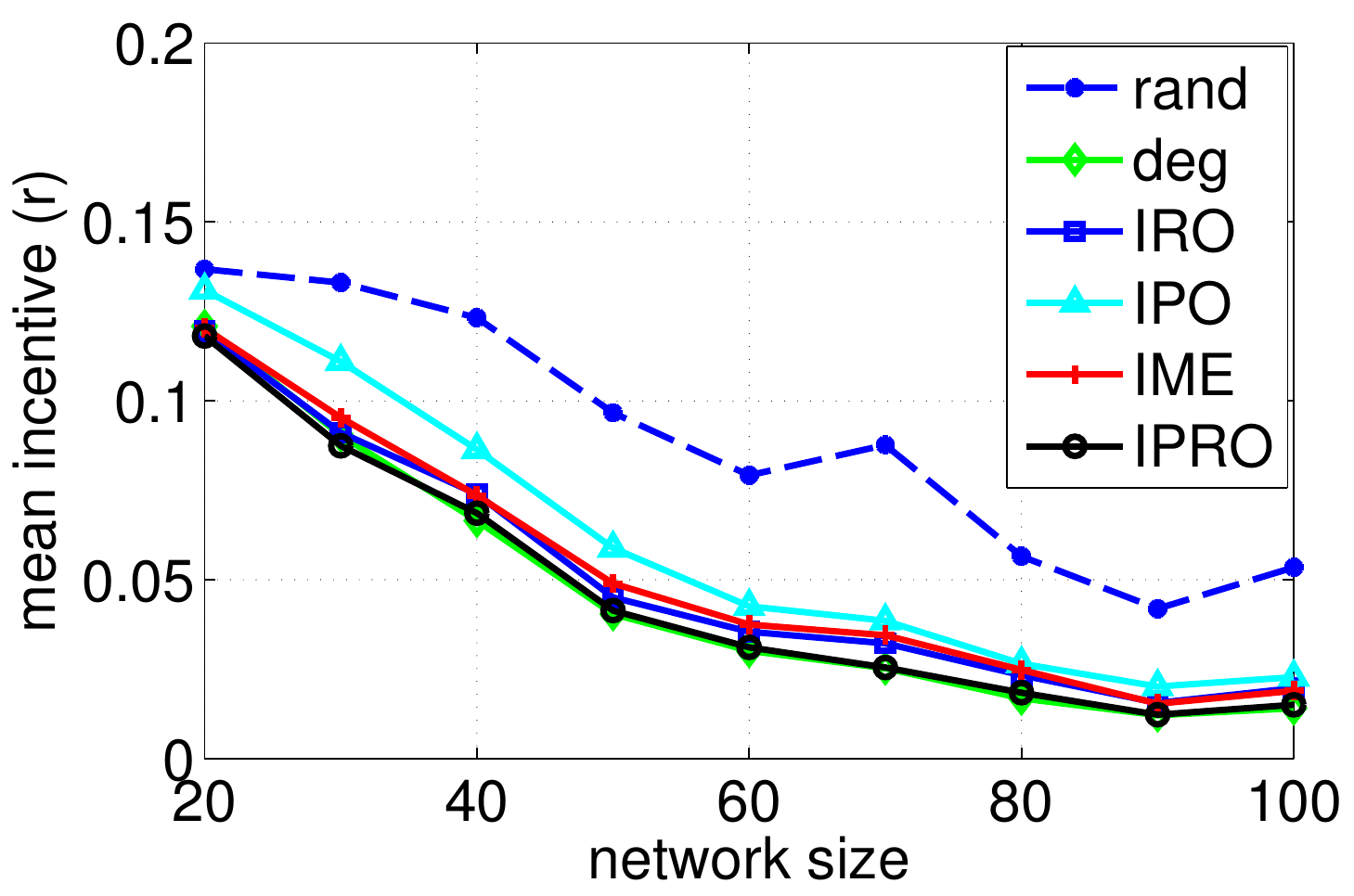}
\caption{Comparison of algorithms for different sizes of networks. The connection radius, threshold distribution, and payoffs are generated exactly as in the simulations for Fig. \ref{figUniformVsTargeted}.}
\label{figPayoffControlUnlimited}
\end{figure}

Next, we compare algorithm performance for various sizes of networks of opponent-coordinating agents, using the same network setup as the previous section.
Fig. \ref{figPayoffControlUnlimited} shows that in this case the IPRO and degree-based algorithms perform the best across all network sizes.

\subsection{Targeted-Reward Control: Network Connectivity}

\begin{figure}[h] 
\centering
\includegraphics[width=.8\linewidth]{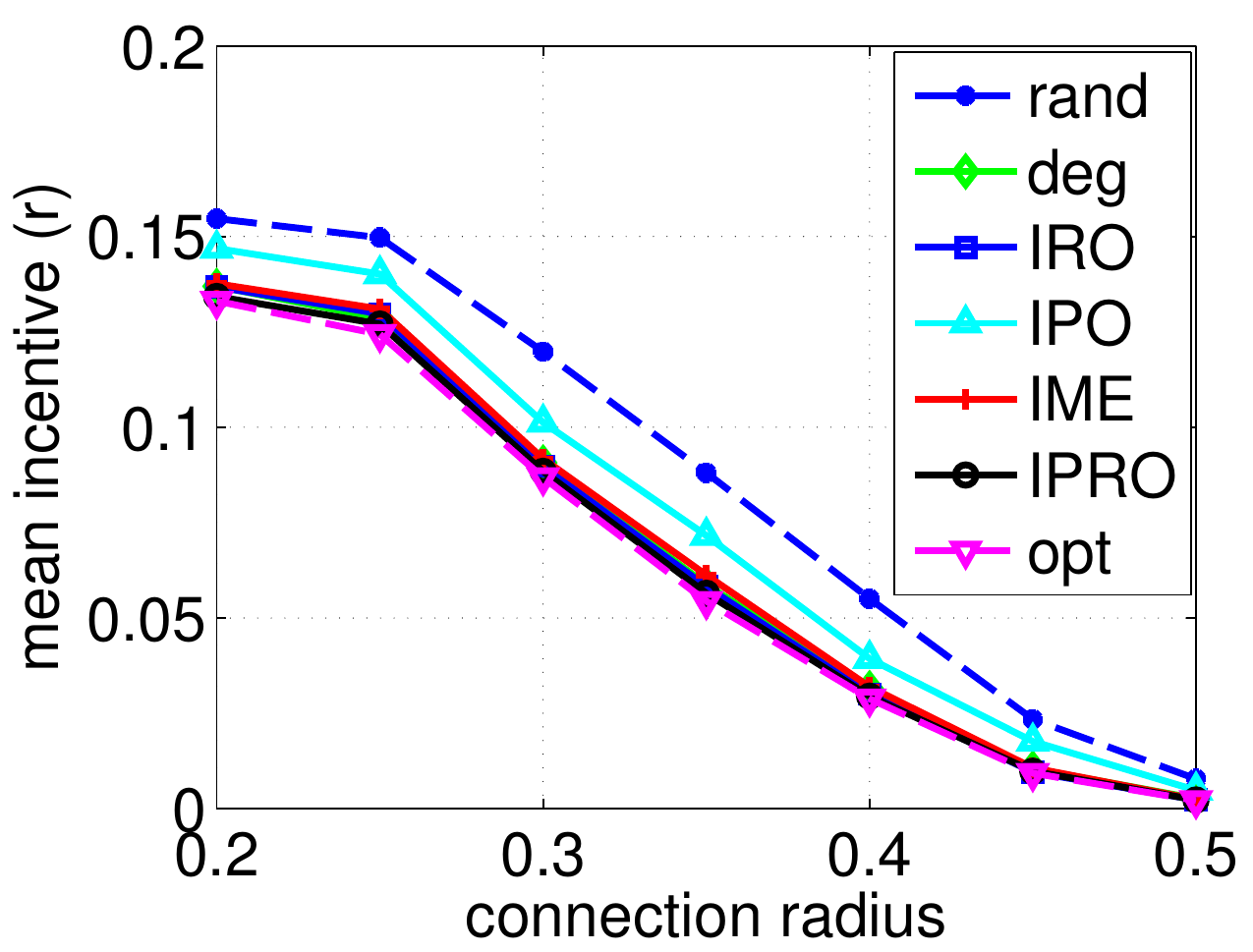}
\caption{Comparison of algorithms on 500 sparsely to densely connected 20-node geometric random networks.}
\label{figConnectivity}
\end{figure}

We now investigate the effect of network connectivity on the total reward required to achieve consensus in strategy $A$. We consider geometric random networks of 20 agents, which is small enough that we can compute the optimal solution using an exhaustive search algorithm and compare this with the proposed algorithms. Fig. \ref{figConnectivity} shows that there is a sharp decrease in the mean incentive required as the networks become more densely connected, likely due to the fact that high-earning agents become more influential. All of the algorithms except for random and IPO yielded near-optimal results in these tests, with IPRO performing the best.
%


\subsection{Targeted-Reward Control: Payoff Variance}

\begin{figure}[h] 
\centering
\includegraphics[width=.8\linewidth]{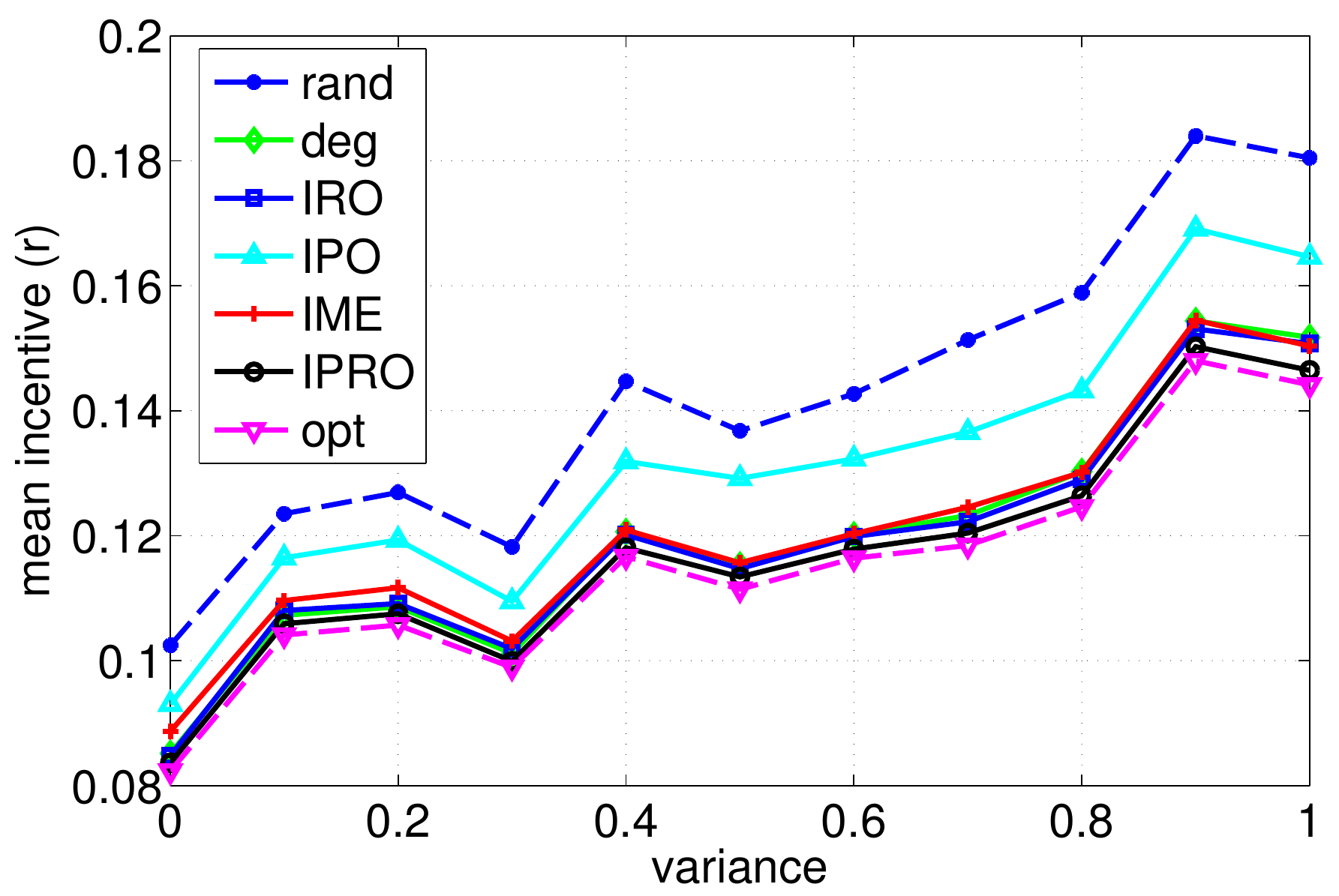}
\caption{Comparison of algorithms for different amounts of payoff variance $v_i$. 500 20-node networks are tested for each value of $v_i$.}
\label{figThresholdVariance}
\end{figure}

Finally, we vary the parameter $v_i$ to understand how the algorithms perform for varying degrees of agent heterogeneity. Fig. \ref{figThresholdVariance} shows that the IPRO algorithm performs the best of the algorithms regardless of the degree of homogeneity or heterogeneity of the agents.

\begin{table}[h]
\def\arraystretch{1.2}
\setlength\tabcolsep{5pt}
\begin{center}
\caption{Simulation Results: Mean Incentives} \label{SummaryTable}
\begin{tabular}{c c c c c} 
\hline 
Algorithm & Fig. 2 & Fig. 3 & Fig. 4  \\  [1ex]  
\hline
rand & .090 & .086  & .143 \\ 
deg & \textbf{.046} & .066 & .120 \\ 
IRO & .051 & .065 & .119 \\
IPO & .060 & .075 & .131 \\
IME & .052 & .067 & .121 \\
\textbf{IPRO} & \textbf{.046} & \textbf{.064} & \textbf{.117} \\
\textcolor{blue}{opt} & -- & \textcolor{blue}{.063} & \textcolor{blue}{.116} \\
\hline
\end{tabular}
\end{center}
\end{table}

\section{Concluding Remarks}
We have revealed three properties of asynchronous \textit{$A$-coordinating} network games under any update rule after incentives are offered to agents when the network is at equilibrium: (i) no agent will switch from $A$ to $B$; (ii) switches occur independent of the sequence in which agents activate; (iii) the network will converge to a unique equilibrium.
This predictability after offering rewards facilitates the design of efficient and in some cases optimal control protocols.
We have further shown that a subset of networks in which agents asynchronously imitate their highest earning neighbor, i.e., networks of opponent-coordinating agents, are indeed $A$-coordinating, and therefore satisfy the above three properties.
Based on this result, we proposed protocols for three control problems that apply to this class of networks: uniform reward control, targeted reward control, and budgeted targeted reward control.
In particular, the IPRO algorithm, which iteratively chooses agents who maximize the ratio of change in potential to offered reward, performs near-optimal in several different cases and outperforms those based on other heuristics such as maximum payoff-earning or minimum required reward. 
In the future, it would be useful to extend this research to networks containing non-coordinating agents.
Although a similar approach may remain effective for this more general case, it might require relaxing the problem statements since it is unlikely that monotonicity and unique convergence will still hold.


\bibliographystyle{IEEEtran}
\bibliography{bib}

\end{document}

%% file: headers.tex
\usepackage{amsfonts}
\usepackage{amsmath}
\usepackage{mathtools}
\usepackage{amssymb}
\usepackage{setspace}
\usepackage{datetime}
\usepackage{fullpage}
\usepackage{graphicx}
\usepackage{algorithm2e}
\usepackage{subfigure}
\usepackage{amsthm}
\usepackage{color}
\usepackage[usenames,dvipsnames]{xcolor}
\usepackage[noadjust]{cite}
\usepackage[usenames,dvipsnames]{xcolor}
\usepackage{tikz}
\usetikzlibrary{calc}
\usetikzlibrary{decorations.markings}
\tikzset{
	->-/.style={decoration={
  		markings,
  		mark=at position #1 with {\arrow{>}}},postaction={decorate}}
  	}
\tikzset{
	-<-/.style={decoration={
  		markings,
  		mark=at position #1 with {\arrow{<}}},postaction={decorate}}

  	  	} 

\newtheorem{theorem}{Theorem}
\newtheorem{lemma}{Lemma}
\newtheorem{corollary}{Corollary}
\newtheorem{problem}{Problem}
\newtheorem{proposition}{Proposition}
\newtheorem{assumption}{Assumption}
\newtheorem{definition}{Definition}

\newcommand{\A}{\mathcal{A}}
\newcommand{\R}{\mathcal{R}}
\newcommand{\N}{\mathcal{N}}
\renewcommand{\S}{\mathcal{S}}
\newcommand{\G}{\mathbb{G}}

\newcommand{\V}{\mathcal{V}}
\newcommand{\E}{\mathcal{E}}

\newcommand{\Rr}{\mathfrak{R}}
\newcommand{\I}{\mathfrak{I}}

\def\str#1{{\mathcal #1}}
\DeclarePairedDelimiter{\ceil}{\lceil}{\rceil}

\newcommand{\dfb}{\stackrel{\Delta}{=}}

%% file: mainLong.bbl
\begin{thebibliography}{10}
\providecommand{\url}[1]{#1}
\csname url@samestyle\endcsname
\providecommand{\newblock}{\relax}
\providecommand{\bibinfo}[2]{#2}
\providecommand{\BIBentrySTDinterwordspacing}{\spaceskip=0pt\relax}
\providecommand{\BIBentryALTinterwordstretchfactor}{4}
\providecommand{\BIBentryALTinterwordspacing}{\spaceskip=\fontdimen2\font plus
\BIBentryALTinterwordstretchfactor\fontdimen3\font minus
  \fontdimen4\font\relax}
\providecommand{\BIBforeignlanguage}[2]{{%
\expandafter\ifx\csname l@#1\endcsname\relax
\typeout{** WARNING: IEEEtran.bst: No hyphenation pattern has been}%
\typeout{** loaded for the language `#1'. Using the pattern for}%
\typeout{** the default language instead.}%
\else
\language=\csname l@#1\endcsname
\fi
#2}}
\providecommand{\BIBdecl}{\relax}
\BIBdecl

\bibitem{traulsen2010human}
A.~Traulsen, D.~Semmann, R.~D. Sommerfeld, H.-J. Krambeck, and M.~Milinski,
  ``Human strategy updating in evolutionary games,'' \emph{Proceedings of the
  National Academy of Sciences}, vol. 107, no.~7, pp. 2962--2966, 2010.

\bibitem{van2015focus}
P.~van~den Berg, L.~Molleman, and F.~J. Weissing, ``Focus on the success of
  others leads to selfish behavior,'' \emph{Proceedings of the National Academy
  of Sciences}, vol. 112, no.~9, pp. 2912--2917, 2015.

\bibitem{ramazi2017imitation}
P.~Ramazi, J.~Riehl, and M.~Cao, ``Imitating successful neighbors hinders
  reaching satisfactory decisions,'' \emph{under review}, 2017.

\bibitem{cortes2015self}
A.~Cort{\'e}s and S.~Martinez, ``Self-triggered best-response dynamics for
  continuous games,'' \emph{IEEE Transactions on Automatic Control}, vol.~60,
  no.~4, pp. 1115--1120, 2015.

\bibitem{marden2014generalized}
J.~R. Marden and T.~Roughgarden, ``Generalized efficiency bounds in distributed
  resource allocation,'' \emph{IEEE Transactions on Automatic Control},
  vol.~59, no.~3, pp. 571--584, 2014.

\bibitem{li2014decoupling}
N.~Li and J.~R. Marden, ``Decoupling coupled constraints through utility
  design,'' \emph{IEEE Transactions on Automatic Control}, vol.~59, no.~8, pp.
  2289--2294, 2014.

\bibitem{wikecek2011stochastic}
P.~Wi{e}cek, E.~Altman, and Y.~Hayel, ``Stochastic state dependent population
  games in wireless communication,'' \emph{IEEE Transactions on Automatic
  Control}, vol.~56, no.~3, pp. 492--505, 2011.

\bibitem{shamma2005dynamic}
J.~S. Shamma and G.~Arslan, ``Dynamic fictitious play, dynamic gradient play,
  and distributed convergence to nash equilibria,'' \emph{IEEE Transactions on
  Automatic Control}, vol.~50, no.~3, pp. 312--327, 2005.

\bibitem{guo2013algebraic}
P.~Guo, Y.~Wang, and H.~Li, ``Algebraic formulation and strategy optimization
  for a class of evolutionary networked games via semi-tensor product method,''
  \emph{Automatica}, vol.~49, no.~11, pp. 3384--3389, 2013.

\bibitem{gharesifard2012evolution}
B.~Gharesifard and J.~Cort{\'e}s, ``Evolution of players' misperceptions in
  hypergames under perfect observations,'' \emph{IEEE Transactions on Automatic
  Control}, vol.~57, no.~7, pp. 1627--1640, 2012.

\bibitem{mylvaganam2015constructive}
T.~Mylvaganam, M.~Sassano, and A.~Astolfi, ``Constructive-nash equilibria for
  nonzero-sum differential games,'' \emph{IEEE Transactions on Automatic
  Control}, vol.~60, no.~4, pp. 950--965, 2015.

\bibitem{cheng2015modeling}
D.~Cheng, F.~He, H.~Qi, and T.~Xu, ``Modeling, analysis and control of
  networked evolutionary games,'' \emph{Automatic Control, IEEE Transactions
  on}, vol.~60, no.~9, pp. 2402--2415, 2015.

\bibitem{ramazi2016networks}
P.~Ramazi, J.~Riehl, and M.~Cao, ``Networks of all confirming and all
  non-confirming individuals tend to reach satisfactory decisions,''
  \emph{Proceedings of the National Academy of Sciences}, 2016.

\bibitem{nowak2006evolutionary}
M.~A. Nowak, \emph{Evolutionary Dynamics: Exploring the Equations of
  Life}.\hskip 1em plus 0.5em minus 0.4em\relax Harvard University Press, 2006.

\bibitem{zhu2016evolutionary}
B.~Zhu, X.~Xia, and Z.~Wu, ``Evolutionary game theoretic demand-side management
  and control for a class of networked smart grid,'' \emph{Automatica},
  vol.~70, pp. 94--100, 2016.

\bibitem{ramazi2016evolutionary}
P.~Ramazi, M.~Cao, and F.~J. Weissing, ``Evolutionary dynamics of homophily and
  heterophily,'' \emph{Scientific reports}, vol.~6, 2016.

\bibitem{ramazi2015feeling}
P.~Ramazi, J.~Hessel, and M.~Cao, ``How feeling betrayed affects cooperation,''
  \emph{PloS one}, vol.~10, no.~4, p. e0122205, 2015.

\bibitem{montanari2010spread}
A.~Montanari and A.~Saberi, ``The spread of innovations in social networks,''
  \emph{Proceedings of the National Academy of Sciences}, vol. 107, no.~47, pp.
  20\,196--20\,201, 2010.

\bibitem{JamesMing2}
J.~R. Riehl and M.~Cao, ``Towards optimal control of evolutionary games on
  networks,'' \emph{IEEE Transactions on Automatic Control}, vol.~62, no.~1,
  pp. 458--462, 2017.

\bibitem{ramazi2015analysis}
P.~Ramazi and M.~Cao, ``Analysis and control of strategic interactions in
  finite heterogeneous populations under best-response update rule,'' in
  \emph{2015 54th IEEE Conference on Decision and Control (CDC)}.\hskip 1em
  plus 0.5em minus 0.4em\relax IEEE, 2015, pp. 4537--4542.

\bibitem{Liang2015}
H.~Liang, M.~Cao, and X.~Wang, ``Analysis and shifting of stochastically stable
  equilibria for evolutionary snowdrift games,'' \emph{Systems \& Control
  Letters}, vol.~85, no.~3, pp. 16--22, 2015.

\bibitem{riehlControl}
J.~Riehl, P.~Ramazi, and M.~Cao, ``Control of asynchronous best-response
  dynamics on networks through payoff incentives,'' \emph{Automatica (under
  review)}.

\bibitem{james}
J.~Riehl and M.~Cao, ``Towards control of evolutionary games on networks,''
  \emph{In Proc. of the 53rd IEEE Conference on Decision and Control}, pp.
  2877--2882, 2014.

\end{thebibliography}
